\documentstyle[aps]{revtex}

\input{psfig}
\newcommand{\vs}{\vspace*}
\begin{document}
\draft
\preprint{}

\title{\bf Scaling violations: Connections between elastic and 
inelastic hadron scattering in a geometrical approach}

\author{P.C. Beggio and M.J. Menon} 

\address{Instituto de F\'{\i}sica ``Gleb Wataghin'',\\
Universidade Estadual de Campinas, Unicamp,  \\
13083-970 Campinas, SP, Brazil}

\author {P. Valin}

\address{D\'ept. de Physique, Universit\'e de Montr\'eal, \\
CP. 6128, Succ. Centre-Ville\\
Montr\'eal, Qu\'ebec, Canada H3C 3J7}


\maketitle

\begin{abstract}
Starting from a short range expansion of the inelastic overlap 
function, capable of describing quite well the elastic $pp$ and
$\overline{p}p$ scattering data, we obtain extensions to the 
inelastic channel, through unitarity and an impact parameter 
approach. 
Based on geometrical arguments
we infer some characteristics of the elementary hadronic process 
and this allows an excellent description of the inclusive 
multiplicity distributions in $pp$ and $\overline{p}p$ collisions. 
With this approach we quantitatively correlate the violations of 
both geometrical and KNO scaling in an analytical way. The physical 
picture from both channels is that the geometrical evolution of the 
hadronic constituents is principally reponsible for the energy 
dependence of the 
physical quantities rather than the dynamical (elementary) interaction 
itself.

\end{abstract}

\vskip 0.5truecm


e-mail: menon@ifi.unicamp.br, fax: 55-19-7885512, phone: 
(19) 7885530

\vskip 1.0truecm

\centerline{\bf Table of Contents}

\vskip 0.5truecm

\leftline{\bf I. INTRODUCTION}

\vskip 0.5truecm

\leftline{\bf II. EXPERIMENTAL DATA AND PHENOMENOLOGICAL CONTEXT }
\vskip 0.2truecm

\leftline{\bf A. Elastic channel}
\leftline{\bf B. Inelastic channel}
\leftline{\bf C. Strategies}

\vskip 0.5truecm

\leftline{\bf III. UNITARITY AND IMPACT PARAMETER PICTURE}
\vskip 0.2truecm

\leftline{\bf A. Hadronic and elementary multiplicity 
distributions}
\leftline{\bf B. Elastic channel input: the BEL $G_{in}$}
\leftline{\bf C. Elementary hadronic process in a geometrical 
picture}

$\bullet$ Analytical relation between multiplicity function and 
eikonal

$\bullet$ Elementary multiplicity distribution

$\bullet$ Power coefficient

\leftline{\bf D. Results for the hadronic multiplicity 
distributions}

\vskip 0.5truecm

\leftline{\bf IV DISCUSSION}
\vskip 0.2truecm

\leftline{\bf A. Sensitivity of the parametrizations}

$\bullet$ Changing $G_{in}$

$\bullet$ Changing the elementary distribution $\varphi$

$\bullet$ Changing the power coefficient $\gamma$

\leftline{\bf B. The multiplicity function and the power 
assumption}

\leftline{\bf C. Physical picture}

\vskip 0.5truecm

\leftline{\bf V. CONCLUSIONS AND FINAL REMARKS}

\newpage

\centerline{\bf I. INTRODUCTION}

\vskip 0.3truecm

Hadron scattering is presently one of the most intriguing process 
in
high energy particle physics. Unlike the unification 
scheme which characterizes the electroweak sector of the standard 
model, some topical
aspects of quantum chromodynamics (QCD) remain yet unknown 
and
this has been a great challenge for decades. One point concerns 
some subtleties emerging from its running coupling constant, which
entails that high energy hadronic phenomena have been 
classified
into two wide and nearly incongrous areas, namely, large $p_T$ 
or hard
processes and low $p_T$ or soft hadronic physics. From a purely 
theoretical
point of view (QCD), these phenomena are treated through 
perturbative and
non-perturbative approaches respectively, and this renders 
difficult an unified formalism able to describe the totality 
of experimental data available on high energy
hadronic interactions. The reason is that, despite the successes 
of perturbative
QCD in the description of the hard (inelastic) hadronic scattering
 and also the
successes of non-perturbative QCD in treating static properties of 
the
hadronic systems, the {\it scattering states in the soft (long 
range) region} 
yet remains without a {\it pure} QCD explanation: Perturbative 
approaches do not
apply and {\it pure} non-perturbative formalisms are not yet able 
to predict
the bulk of the scattering states.

This soft hadronic physics is associated with the elastic and 
diffractive scattering, characterized, experimentally, by several 
physical quantities
in both the elastic and inelastic channels, such as elastic 
differential
cross section, total cross section, charged multiplicity 
distributions, average multiplicities and others \cite{leader}. 
In spite of the
long efforts to describe these data through a pure microscopic 
theory (QCD), our knowledge is still largely phenomenological 
and also based on a wide class of models and some distinct
theoretical concepts, such as Pomeron, Odderon, impact parameter 
picture, parton and dual models, Monte Carlo approaches and so on. 

However, presently, this phenomenological treatment of the soft 
hadronic physics plays an important role as a source of new 
theoretical insights and as
a strategy in the search for adequate calculational schemes in 
QCD. The multiple facets associated with this phenomenological 
scenario have been extensively discussed in the literature and 
Ref. \cite{blois}  presents a detailed outlook of the progresses 
and present status of the area.

In addition to the intrinsic importance of high energy diffractive 
physics associated
with our limited theoretical understanding, a renewed interest in 
the subject may be seen in the last years. This, in part, is due 
to the Hera and Tevatron programs, but also to the advent of the
 next accelerator generation, the 
Relativistic Heavy Ion Collider (RHIC) and the CERN Large Hadron 
Collider (LHC).
In fact, with these new machines it shall be possible to 
investigate $pp$ collisions at center-of-mass energies never 
reached before in accelerators,
allowing comparative studies between $pp$ and $\overline p p$ 
scattering at the highest energies, including both hard and soft 
processes.

Presently, at this ``pre-new-era'' stage and due to the {\it lack 
of an widely
accepted unified theoretical treatment of both elastic and inelastic 
channels}, it may be important to 
re-investigate {\it ways of connecting} these channels, 
looking for
new information. Even if the treatment is essentially 
phenomenological, as
explained before, the predictions shall be checked and may 
contribute with
future theoretical (QCD) developments.

To this end, in this work we shall investigate some aspects of
 both
elastic and inelastic $pp$ and $\overline p p$ scattering in the 
context of a particular phenomenological
approach. Our goal is to obtain simultaneous descriptions of some
 experimental data from both channels, that is, our primary 
interest concerns {\it connections
between elastic and inelastic hadron scattering}. Accordingly we
shall base our
study on one of the most important principle of quantum field 
theory: Unitarity. 

For reasons that will be explained in detail in what follows, our 
framework is the
impact parameter formalism (geometrical approach). At first, under 
geometrical considerations, we shall not refer to quarks/gluons or 
partons, but 
treat
hadron-hadron interactions as collisions between composite objects 
made up by elementary parts, which we shall generically refer as
 ``constituents''. At the
end we discuss some possible connections between our results and 
the framework of QCD. Also, as shall be explained, our starting 
point is the description of physical quantities that characterize 
the elastic channel in $pp$ and $\overline p p$ scattering. We 
then proceed to consider the inelastic channel through unitarity 
arguments and in a geometrical approach.

Following other authors \cite{barshay,ly}, we shall
express the ``complex" (overall) hadron-$p$ multiplicity 
distributions 
(inelastic channel) in terms of an ``elementary" distribution 
(associated with an elementary process taking place at given 
impact 
parameter) and the inelastic overlap function, which is 
constructed 
from descriptions of the elastic channel data. The novel aspects 
concern: (a) quantitative 
correlation between the {\it violations} of the Koba-Nielsen-Olesen 
(KNO) scaling \cite{kno} (inelastic 
channel) and geometrical scaling \cite{deus} (elastic channel); 
(b) introduction 
of novel parametrizations for the elementary quantities based on  
geometrical arguments and taking suitable account
of the most recent data on {\it contact interactions}. 
With this general formalism, in addition to the description of the 
elastic data (even at large momentum transfers), the hadronic 
multiplicity distributions may be 
evaluated 
and an excellent reproduction of the 
experimental data on $pp$ and $\overline{p}p$  
inelastic multiplicities is achieved \cite{xrt}. We also present 
predictions 
at the LHC energies.

The paper is organized as follows. In Sec. II we discuss the 
underlying phenomenological ideas, the data to be investigated and 
the
strategy assumed. In Sec. III we present the theoretical framework
connecting elastic and inelastic channels, the inputs from elastic 
scattering
data, the novel parametrizations for the elementary processes, the 
predictions for the hadronic multiplicities distributions and 
comparison with the experimental data.
In Sec. IV we discuss in some detail all the results obtained and 
the physical and geometrical interpretations. The conclusions and 
some final remarks are the content 
of Sec. V.

In what follows we shall represent the main physical quantities 
associated 
with hadron-hadron scattering (complex/overall system) by capital 
letters and those associated with constituent-constituent 
interactions (elementary process) by lower case.

\vskip 0.3truecm

\centerline{\bf II. EXPERIMENTAL DATA AND PHENOMENOLOGICAL CONTEXT}

\vskip 0.3truecm

The broad classification in hard, soft (and also semi-hard) 
processes is based
on the momentum transferred in the collision. On the other hand, 
depending on the
physical process involved, high energy hadron scattering may also 
be classified
into elastic and inelastic processes and the later, in diffractive 
(single and
double dissociation) and non diffractive. Concerning {\it both 
elastic and
inelastic channels}, one of the striking features that emerged 
from early
experiments was the violation of the scaling laws, namely, the 
geometrical
scaling in elastic scattering \cite{deus} and the KNO scaling in 
the inelastic
events \cite{kno}.
For this reason, our main interest in this work is to correlate 
quantitatively the above
scaling violations and to discuss its phenomenological and 
dynamical aspects.
To this end, before we present the underlying formalism and results, 
we discuss in this section the physical observables to be investigated 
and the reasons for our choices
concerning phenomenology and strategies.

\vskip 0.2truecm
\centerline{\bf A. Elastic channel}
\vskip 0.2truecm

The differential cross section is the most important physical 
observable
in the elastic channel, since from it other quantities may be obtained,
in particular, the integrated elastic cross section, $\sigma_{el}$, and 
the total
cross section, $\sigma_{tot}$ (optical theorem). The violation of the
geometrical scaling may be characterized by the increase of the ratio
$\sigma_{el}/ \sigma_{tot}$ with the energy at the CERN Intersecting 
Storage
Ring ($ ISR $) and at the CERN Super Proton Synchrotron 
($S\overline{p}pS$) regions.

The differential cross section yields the elastic hadronic amplitude,
$F(q,s)$, by
\begin{equation}
{d\sigma \over dq^2} = \pi |F(q,s)|^2
\end{equation}
and this amplitude may be expressed in terms of the elastic profile 
function,
$\Gamma (b,s)$, by

\begin{equation}
F(q,s)=i\int bdbJ_{0}(qb)\Gamma (b,s),
\end{equation}
where $b$ is the impact parameterand $\sqrt s $ the center-of-mass 
energy.

As commented before, despite the bulk of models able to reproduce the 
differential
cross section data at the $ ISR $, $S\overline{p}pS$ and Tevatron 
energies \cite{blois}, an
approach based exclusively in QCD is still missing. Obviously due to 
its {\it soft} character, a
QCD treatment of the elastic scattering should be non-perturbative. 
Along this line,
despite the difficulties, 
important results have recently been reached through the works by 
Landshoff, Nachtmann,
Simonov, Dosch, Ferreira and Kramer \cite{ln,dosch}. The approach, 
based 
on the functional integral representation (QCD)
and eikonal approximation, allows to extract a quark-quark profile 
function 
$\gamma (b)$ (impact parameter space) from the gluon gauge-invariant 
two-point
correlation function, determined, for example, from lattice QCD 
\cite{digi}. Through the Fourier transform (analogous to Eq. (2)
at the elementary level), the quark-quark scattering amplitude, 
$f(q,s)$, 
may be obtained:

\begin{equation}
f(q,s)=i\int bdbJ_{0}(qb)\gamma (b,s).
\end{equation}

One possible connection with the hadronic scattering amplitude, 
Eq. (2), 
is by means of the
Stocastic Vacuum Model (SVM) \cite{dosch} and some important results 
have 
recently been obtained \cite{qcddata}. However, presently, this 
theoretical 
framework still
depends on some phenomenological inputs. Also, it  is able to
reproduce only the experimental data in the forward region and/or 
very 
small values of the momentum transfer and does not distinguish $pp$ 
and 
$\overline{p}p$ scattering
(dip region), even at $ ISR $ energies \cite{qcddata}.

Another way to obtain the hadronic amplitude from the elementary one 
is
through the Glauber's multiple diffraction theory (MDT) \cite{glauber} 
and this plays a central role in our choices concerning phenomenology 
and strategies as discussed in what follows.

Originally the MDT was applied to hadron-nucleus and nucleus-nucleus 
collisions \cite{glauber} and
after to hadron-hadron scattering \cite{gv}. The topical point which 
interest us is the allowed
general connection between complex quantity (composite object) and 
elementary quantity
(constituents). In the case of hadron-hadron collisions, the 
connection 
between the
hadronic amplitude (composite object) and the elementary amplitude 
(constituents)
is also established through the eikonal approximation. In this 
approach 
the
hadronic profile function, Eq. (2), is expressed by

\begin{equation}
\Gamma (b,s) = 1-e^{i\chi (b,s)},
\end{equation}
where

\begin{equation}
\chi (b,s)= C\int qdqJ_{0}(qb)G_{A}G_{B}f
\end{equation}
is the eikonal function, $G_{A,B}$ the hadronic form
factors, $f$ the elementary (constituent - constituent) amplitude
and $C$ does not depend on the transferred momentum. The above notation 
\cite{menon} shall be useful for the discussion we are interested in.

In spite of their simplicity, Eqs. (2), (4) and (5) are extremely 
useful. 
Recently, with
suitable parametrizations for the form factors and with the elementary 
amplitude
(quark-quark) extracted from a parametrization for the gluonic 
correlator 
through the functional
approach (non-perturbative QCD), Grandel and Weise obtained good 
descriptions of the
differential cross secion data for $pp$ and $\overline{p}p$  elastic 
scattering at the $ ISR $ and $S\overline{p}pS$ energies, but only in 
the region of small momentum
transfer \cite{gw}. On the other hand, excellent descriptions of 
experimental 
data, including also large
momentum transfers, have been obtained in a rather phenomenological 
context, through
suitable parametrizations for $G_A$, $G_B$ and $f$ \cite{mdmodels}. 
Moreover, elementary
amplitudes obtained through the SVM and the gluonic correlator from 
lattice QCD have been
investigated and also comparisons with empirical analysis and model 
predictions have been discussed \cite{menon,mmt}.

We understand that all these facts indicate that the impact parameter 
formalism
(and the eikonal approximation), connecting the complex (overall) 
amplitude with the
elementary amplitude (constituent-constituent), Eqs. (2), (4) and (5), 
is a very fruitful and
simple approach in the investigation of the elastic hadron scattering. 
As shown, it
seems also to be an adequate bridge between phenomenology and 
non-perturbative
QCD. These conclusions constitute one of the foundations of our 
approach 
and,
as discussed in what follows, the {\it extensions to the inelastic 
channel shall be
based on the general idea of connections between overall and 
elementary 
quantities in an impact parameter picture}.

\vskip 0.2truecm
\centerline{\bf B. Inelastic channel}
\vskip 0.2truecm

Concerning scaling in the inelastic channel, the quantity of interest 
is the
hadronic charged particle multiplicity distribution $P_N$, normalized 
in terms of
the KNO variable, $Z = N(s)/<N>(s)$, as

\begin{equation}
<N>(s) P_N(Z) \equiv \Phi.
\end{equation}
The broader distribution observed at the $S\overline{p}pS$ 
characterizes 
the
violation of the KNO scaling, namely, $\Phi = \Phi (Z,s)$.

As in the case of elastic differential cross section data, a wide 
class 
of
models describes this behavior, as for example, dual parton 
\cite{dual}, 
fireball
\cite{fireball}, two-component  models \cite{2comp}, and others. Also, 
hadronic processes have been extensively treated through Monte Carlo 
event generators and
the Lund parton approach \cite{lund,herwig,ariadne}. However, we 
observe 
that,
despite some QCD inspired approaches and good descriptions of some 
soft 
processes,
all these formalims and models are concerned {\it exclusively} with 
the 
inelastic 
channel and this
is the topical point that distinguishes our strategy, as discussed 
in what 
follows. We shall also return to this subject in Secs. IV and V.

\vskip 0.2truecm
\centerline{\bf C. Strategies}
\vskip 0.2truecm

Connections between Geometrical and KNO scalings were established a 
long time
ago, by Dias de Deus \cite{deus}, Lam and Yeung \cite{ly}. However, 
we are interested here in
their violations and the central point is: Does one need a new 
connection 
when the two phenomena are violated at the $Sp\overline{p}S$  or 
can the 
two effects be correlated both phenomenologically and dynamically? We 
will argue that the latter alternative seems to be prevail. 
Specifically, 
our goal is to correlate quantitatively both
violations in an analytical way and we shall show that, beginning with 
a formalism that
describes quite well the violation of the Geometrical scaling (elastic 
channel input),
it is possible to extend it and to describe, quantitatively, the 
violation 
of the KNO scaling in
an analytical way. We stress that this strategy distinguishes our 
approach from all the other model and theoretical descriptions of 
elastic or inelastic scattering that
treat these interactions separately, in an independent way or in 
distinct contexts.

Since the connection between {\it elastic channel $\rightarrow$ 
inelastic channel} 
is our primary interest, the
approach shall be based in direct analogy with the ideas discussed 
in
Sec. II.A, that is, we consider hadron-hadron collisions as collisions 
between  complex objects, each one composed by a number of more 
elementary 
ones. As an extension of the Glauber multiple diffraction theory, 
which 
connects hadronic and elementary elastic amplitudes, 
we shall consider the impact parameter formalism and also
express the hadronic multiplicity distribution (complex system) in 
terms of elementary
multiplicity distributions (constituents). The point is to describe 
a ``wide'' distribution (hadronic) by superimposing a number of 
narrower 
ones (elementary) \cite{ly}.
What we shall do here is to infer what these elementary distributions 
should
be, in order to reproduce the experimental data on hadronic 
distributions 
and 
in the context of the impact parameter picture. He hope that, as in 
the 
elastic
case, these information can contribute to further theoretical 
developments.

\vskip 0.3truecm

\centerline{\bf III. UNITARITY AND IMPACT PARAMETER PICTURE}

\vskip 0.3truecm

Unitarity is one of the most important principles in quantum field 
theory.
In the geometrical picture, unitarity correlates the elastic 
scattering 
amplitude in the impact parameter $b$ space, $\Gamma(b,s)$, Eq. (2), 
with 
the inelastic overlap function, $G_{in}(b,s)$, by

\begin{equation}
2 Re \Gamma(b,s)=|\Gamma(b,s)|^{2}+G_{in}(b,s)
\end{equation}
which is term-by-term equivalent to \cite{hv1} 

\begin{equation}
G_{tot}(b,s)=G_{el}(b,s)+G_{in}(b,s).
\end{equation}
For a purely imaginary elastic amplitude in momentum transfer space 
the profile function $\Gamma(b,s)$ is real and in the eikonal 
approximation is expressed by

\begin{equation} 
\Gamma(b,s)=1-exp[-\Omega(b,s)],
\end{equation}
where $\Omega(b,s) = Im\chi (b,s)$ in Eq. (4). With this,

\begin{equation} 
G_{in}(b,s)=1-exp[-2\ \Omega(b,s)] \equiv \sigma_{in}(b,s)
\end{equation}
is the probability for an inelastic event to take place at $b$ and 
$s$ and

\begin{equation} 
\sigma_{in}(s)=\int d^{2}{\bf b}\ G_{in}(b,s).
\end{equation}

In this picture the topological cross section for producting an even 
number $N$ of charged particles at CM energy $\sqrt{s}$ is given by 
  
\begin{equation}
\sigma_{N}(s)= \int d^{2}{\bf b}\ \sigma_{N}(b,s) =  \int d^{2}
{\bf b}\ 
\sigma_{in}(b,s) \left[ \frac{\sigma_{N}(b,s)}{\sigma_{in}(b,s)} 
\right]
\end{equation} 
where the quantity in brackets can be interpreted as the probability 
of producing $N$ particles at impact parameter $b$.

\vskip0.2truecm
\centerline{\bf A. Hadronic and elementary multiplicity distributions}
\vskip 0.2truecm

We now introduce the multiplicity distributions for both an overall 
and an elementary processes in terms of corresponding KNO variables 
\cite{kno} and also the formal connection between these distributions.

Representing the hadronic (overall) multiplicity distribution by 
$\Phi$ and the 
corresponding KNO variable by 

\begin{equation} 
Z=\frac{N(s)}{<N>(s)}
\end{equation}
where $<N>(s)$ is the average hadronic multiplicity at $\sqrt{s}$, 
we have in 
general 

\begin{equation} 
\Phi=<N>(s) \frac{\sigma_{N}(s)}{\sigma_{in}(s)} = \Phi (Z,s).
\end{equation}

Now, let $<n>(b,s)$ be the average number of particles produced at $b$ 
and $s$, $\varphi$ the elementary multiplicity distribution and 

\begin{equation} 
z=\frac{N(s)}{<n>(b,s)}
\end{equation}
a KNO variable associated with the elementary process taking place at 
$b$ (and $s$). Then, in general, 

\begin{equation} 
\varphi=<n>(b,s) \frac{\sigma_{N}(b,s)}{\sigma_{in}(b,s)} = \varphi 
(z,s).
\end{equation}

Both distributions are normalized by the usual conditions 
\cite{barshay,ly}

\begin{equation}
\int_{0}^{\infty} \Phi(Z) dZ=2=\int_{0}^{\infty} \Phi(Z) Z dZ.
\end{equation}

\begin{equation}
\int_{0}^{\infty} \varphi(z) dz=2=\int_{0}^{\infty} \varphi(z) z dz,
\end{equation}

The relationship between $\Phi$ and $\varphi$ then follows from Eqs. 
(10-12), (14) and (16):

\begin{equation} 
\Phi=\frac{<N>(s) \int d^{2} {\bf b} \frac{G_{in}(b,s)}{<n>(b,s)} 
\varphi}{\int d^{2} {\bf b}\ G_{in} (b,s)}.
\end{equation}

Now, let us define a {\it multiplicity function} $m(b,s)$ by the
ratio

\begin{equation} 
 m(b,s) = {<n>(b,s) \over <N>(s)},
\end{equation}
so that Eq. (19) becomes

\begin{equation} 
\Phi=\frac{\int d^{2} {\bf b} \frac{G_{in}(b,s)}{m(b,s)} \varphi 
(\frac{Z}{m(b,s)})}{ \int d^{2} {\bf b}\  G_{in} (b,s)} = \Phi (Z,s).
\end{equation}

It is well known that connections between KNO and Geometrical scaling 
may be established if $m(b,s)=m(b/R(s))$ and also $G_{in}(b,s)=G_{in}
(b/R(s))$ \cite{deus}, where $R(s)$ is the ``geometrical radius''.
In this case $\Phi (Z,s)$ is only a function of $Z$.

The general result (21) means that, once one has parametrizations 
for $G_{in}(b,s)$ and the elementary quantities $\varphi$ 
(multiplicities distribution) and $m(b,s)$ (multiplicity function) 
the overall hadronic multiplicity distribution may be evaluated. 
In this work we consider $G_{in}(b,s)$ from analyses of elastic $pp$ 
and $\overline{p}p$ scattering data (taking account of geometrical 
scaling violation) and infer the elementary quantities based on 
geometric arguments, as explained in what follows. In so doing, we 
shall correlate quantitatively the {\it violations} of both KNO and 
Geometrical scaling in an analytical way.

\vskip 0.2truecm
\centerline{\bf B. Elastic channel input: the BEL $G_{in}$}
\vskip 0.2truecm

In the elastic channel, the breaking of Geometrical scaling is 
quite well described by the BEL behaviour, analytically expressed
 by the Short Range Expansion of the inelastic overlap function 
\cite{hv1,hv2}

\begin{equation}
G_{in}(b,s)=P(s)exp\{-b^{2}/4B(s)\}k(x,s),
\end{equation} 
with $k$ being expanded in terms of a short-range variable $x=b\ 
exp\{-(\epsilon b)^{2}/4 B(s)\}$, i.e.

\begin{equation}
k(x,s)=\sum_{n=0}^{N} \delta_{2n}(s) \left[ \frac{\epsilon\ 
exp\{1/2\}}{\sqrt{2B(s)}}\ x \right]^{2n}.
\end{equation}

The quantity in the bracket of (23) by itself exhibits GS for 
constant values of $\epsilon^{2}$ $\approx$ 0.78, but $k(x,s)$ 
doesn't because of the $s$-dependence of $\delta_{2n}(s)$ and therefore 
$G_{in}(b,s)$ doesn't either (also because of P($s$)). Each term in the 
bracket of (23) has a maximum value of 1 and the rapid convergence of 
the series reproduces data for all values of $-t$ $\in$ (0, 14) GeV$^2$ 
with N=3. For $k$=1=$P$, we recover the Van Hove limit for $\sigma_{el}/\sigma_{tot}$=1-$\frac{1}{4(1-{\rm ln}2)}$ $\approx$ 
0.1853 which is nearly attained at the ISR. The deviation of $k$ from 
the constant value of 1, in particular the increase of $\delta_{2}(s)$ 
with increasing $s$ is responsible for the {\it Edgier} behaviour of 
$G_{in}(b,s)$, while increasing values of $P(s)$ and $B(s)$ make the 
proton {\it Blacker} and {\it Larger} respectively (BEL behavior of 
the inelastic overlap function).
 
Excellent agreement with experimental data on $pp$ and 
$\overline{p}p$ elastic scattering is achieved \cite{hv2} for the 
following parametrizations in terms of the Froissart-like variable 
$y={\rm ln}^{2}(s/s_{0})$ with $s_{0}=100$ GeV$^2$ 

\begin{equation}
P(s)=\frac{0.908 + 0.027 y}{1 + 0.027 y}, \quad \delta_{2}(s)=
0.115 + 0.00094 y
\end{equation} 

\centerline{$B(s)=6.64 + 0.044 y$}

\leftline{and $\delta_{4}$ determined from $\delta_{2}$ in 
some models \cite{hv3} by $\delta_{4}=\delta_{2}^{2}/4$.}

\vskip 0.2truecm
\centerline{\bf C. Elementary hadronic process in a geometrical
 picture}
\vskip 0.2truecm

We now turn to the discussion of the {\it elementary hadronic process}, 
characterized by $\varphi $ and $m$ in Eq. (21). By construction, these
quantities are associated to collisions of strongly interacting hadronic 
constituents. As commented before, due to the success of the geometrical
 models 
in the investigation of {\it elastic} hadron scattering (for example, 
the 
above BEL approach) and the presently lack of a {\it pure} QCD approach 
and/or 
Monte Carlo models to the subject (elastic/soft scattering states), we 
shall 
discuss what an {\it elementary hadronic} process could be in the 
geometrical 
framework and in an analytical way. Our arguments are as follows.

In the geometrical approach an elementary process is a process occuring 
at a 
given impact parameter. Concerning {\it contact interactions}, 
experimental 
information is {\it only available from lepton-lepton collisions}, 
which is 
a process occuring in a unique angular momentum state and therefore 
also at 
a given impact parameter (zero in this case). Although {\it these 
processes can 
not be the same as collisions between hadrons constituents}, it is 
reasonable, 
from the geometrical point of view, to think that {\it some 
characteristics of both 
processes could be similar}. The point is to find out or infer what 
they 
could be.

For these reasons we shall consider the experimental data available on
$e^+e^-$ collisions as a possible source of (limited) geometrical 
information
concerning elementary hadronic interactions (at given impact parameter).
We do not pretend to look for connections between $e^+e^-$ 
annihilations and 
$pp$ and $\overline pp$ collisions but to extract from the former 
processes 
suitable information that allows the construction of the hadronic 
multiplicities (and the connections with the corresponding elastic 
amplitude) 
in an {\it analytical way and in the geometrical context}. This may 
be 
achieved for both $\varphi$ and $m$ in Eq. (21) through the following
procedure.

\vskip 0.2truecm

$\bullet$ Analytical relation between multiplicity function and eikonal

\vskip 0.2truecm

First, in order to connect the multiplicity function $m(b,s)$ and the 
eikonal
$\Omega (b,s)$ \cite{barshay,ly} (and so $G_{in}(b,s)$ by Eq. (10)) in 
an {\it analytical 
way}, let us consider the very simple assumption that the {\it average
 multiplicity at given impact parameter} 
depends on the center-of-mass energy in the form of a general power law

\begin{equation}
<n>(b\ fixed,s) \propto E_{CM}^{\gamma}.
\end{equation}
We shall discuss this assumption in detail in Sec. IV.B.. 

Now, from  Eqs. (10) 
and (11), $exp\{-2\ \Omega(b,s)\}$ is the transmission coefficient, 
i.e. the probability of having no interaction at a given impact 
parameter, and therefore $\Omega$ should be proportional to the 
thickness of the target, or as first approximation, to the energy 
$E_{CM}$ that can be deposited 
at $b$ for particle production at a given $s$. By Eq. (25) this implies

\begin{equation}
<n>(b,s) \propto  \Omega^{\gamma}(b,s).
\end{equation}
Comparison of Eqs. (20) and (26) allows us to correlate the 
multiplicity function $m(b,s)$ with the eikonal through a 
non-factorizing relation (in $b$ and $s$): 

\begin{equation}
m(b,s)=\xi(s) \Omega^{\gamma}(b,s),
\end{equation}
with $\xi (s)$ being determined by the normalization condition of the 
overall multiplicity distribution, Eq. (18). With this, Eq. (21) 
becomes 

\begin{equation} 
\Phi=\frac{\int d^{2} {\bf b}\ \frac{G_{in}(b,s)}{\xi 
\Omega^{\gamma}(b,s)}\ \varphi (\frac{Z}{\xi \Omega^{\gamma}
(b,s)})}{\int d^{2} {\bf b}\ G_{in} (b,s)}
\end{equation}
where

\begin{equation}
\xi(s)=\frac{\int db^{2}G_{in}(b,s)}{\int db^{2}\ G_{in}(b,s) 
\Omega^{\gamma}(b,s)}.
\end{equation}

Once the above {\it analytical connection} is assumed, the elementary 
hadronic
process is now characterized by only two quantities, namely, the
elementary distribution $\varphi$ and the power coefficient $\gamma$.
We proceed with the determination of these quantities through 
quantitative 
analyses of $e^{+}e^{-}$ data and under the following arguments.

\vskip 0.2truecm
\newpage 

$\bullet$ Elementary multiplicity distribution

\vskip 0.2truecm
 
Because the elementary process occurs at a given impact parameter, 
its elementary structure suggests that it should scale in the KNO 
sense. Now, since experimental information on $e^{+}e^{-}$ 
multiplicity 
distributions shows agreement with this scaling \cite{delphi}, we 
shall base our 
parametrization for $\varphi$ just on these data. In particular, it is 
sufficient to assume a gamma distribution (one free parameter), 
normalized according 
to Eq. (17), 

\begin{equation}
\varphi(z)=2 \frac{K^{K}}{\Gamma (K)} z^{K-1} exp\{-Kz\}.
\end{equation}
Fit to the most recent data, covering the interval 22.0 $GeV$ 
$\leq$ $\sqrt{s}$ $\leq$ 161 GeV \cite{delphi,data1} furnished 
$K=10.775$ 
$\pm$ 0.064 with $\chi^{2}/N_{DF}=508/195 = 2.61$ and the result is 
shown in Fig. 1.

Concerning this fit, we verified that data at $29$ and $56$ GeV 
make the highest contributions in terms of $\chi ^2$ values. For 
example, 
if the former data are excluded we obtain $K = 10.62$ and 
$\chi ^2 / N_{DF} 
= 414/181 = 2.29$ and if both sets are excluded then $K = 10.88$ and 
$\chi ^2 / 
N_{DF} = 286/162 = 1.77$. For comparison we recall that the DELPHI fit 
through a 
negative binomial distribution to data at only $91$ GeV gives 
$\chi ^2 / N_{DF} = 80/34 = 2.35$ (and $\chi ^2 / 
N_{DF} = 43/33 = 1.30$ 
through a modified negative binomial distribution) \cite{delphi}. 
However, we also verified that the above two values for $K$ are not 
so sensitive 
in the final result concerning the hadronic multiplicity distribution, 
which is 
our goal (we shall return to this point in Sec. IV.A). For this reason 
and since we are only looking for experimental 
information that could represent contact interactions (geometrical 
point 
of view) we consider our first result shown in Fig. 1 as the 
representative 
one.

\vskip 0.2truecm

$\bullet$ Power coefficient

\vskip 0.2truecm

Finally, following Eq. (25), we consider fits to the $e^{+}e^{-}$ 
average multiplicity through the general power law

\begin{equation}
<n>_{e^{+}e^{-}}= A [\sqrt{s}]^{\gamma}.
\end{equation}
We collected experimental data at center-of-mass energies above 
resonances and thresholds and also the most recent data at the 
highest energies, covering the interval 5.1 GeV $\leq$ $\sqrt{s}$ 
$\leq$ 183 GeV \cite{data1,data2}. Fitting to Eq. (31) yields $A$=2.09 
$\pm$ 0.02,

\begin{equation}
\gamma=0.516 \pm 0.002
\end{equation}
with $\chi^{2}/N_{DF}=409/46 = 8.89$ and the result is shown in Fig. 2.
We observe that this parametrization deviates from the data 
above $\sqrt{s}$ $\sim$ 100 GeV and this contributes to the high 
$\chi^{2}$ value. However, as commented before, we do not expect 
that $e^{+}e^{-}$ annihilation exactly represent the collisions 
between hadrons constituents. The power law is a form which allows 
an {\it analytical} and simple
connection between the multiplicity function and the eikonal 
as expressed by Eq. (27). In Sec. IV.B we present a detailed discussion 
concerning this power assumption and in Sec. IV.A and IV.C, we discuss 
the physical meaning of the differences between our parametrization and 
the experimental data on $<n>_{e^{+}e^{-}}$.

\vskip 0.2truecm
\centerline{\bf D. Results for the hadronic multiplicities 
distributions}
\vskip 0.2truecm

With the above results we are now able to predict the hadronic 
inelastic multiplicity distribution $\Phi(Z,s)$, Eqs. (28, 29), 
without free 
parameters: $G_{in}(b,s)$ (and $\Omega(b,s)$) comes from analysis 
of the elastic scattering data (Eq. (10) and (22-24)); $\varphi(z)$ 
and $\gamma$ from fits to $e^{+}e^{-}$ data, Eq. (30) and (31-32) 
respectively. 

We express $\Phi$ in terms of the scaling variable 
$Z^{'}=N^{'}/<N^{'}>$ where $<N^{'}>=N(s)-N_{0}$ with $N_{0}$=0.9 
leading charges removed. It is well known \cite{ansorge} that such a 
subtraction improve the KNO curves for all measured data below 
the $S\overline{p}pS$ Collider with the above value of $N_{0}$. 
This is completely equivalent to the Wroblewski \cite{w} relation 
for the dispersion $D=\sqrt{N^{2}-<N>^{2}}$=0.594 $[<N>-N_{0}]$ 
with the same value of $N_{0}$. Values of $N_{0}$ around 1 and
 the numerical value of the $D$ vs $N$ can be found by parton
 model arguments \cite{rudaz} for valence quark distributions.

The predictions for $pp$ scattering at ISR energies and 
$\overline{p}p$ at 546 GeV are shown in Figs. (3) and (4), 
respectively, together with the experimental data \cite{break,ua5}. 
The theoretical curves present excellent agreement with all 
the data, showing a slow evolution with the energy at ISR 
for large $Z^{'}$ and reproducting the KNO violations for large 
$Z^{'}$ values at 546 GeV. In Fig. (4) is also shown the predictions 
at 14 TeV (LHC).

\vskip 0.3truecm

\centerline{\bf IV DISCUSSION}

\vskip 0.3truecm

In the last section we obtained a quantitative correlation between the
violations of both KNO and geometrical scaling. In the framework
of the impact parameter picture, Sec. III.A, we only used four
inputs, three parametrizations from fits to experimental data and one
geometrical assumption, namely,

\begin{enumerate}

\item $G_{in}(b,s)$ from fits to elastic scattering data (BEL 
behavior);

\item $\varphi(z)$ from fit to $e^+ e^-$ multiplicity distribution 
data;

\item The geometrical assumption (27) concerning the multiplicity 
function
$m(b,s)$;

\item  The power coeficient $\gamma$, from fit to $e^+ e^-$ average
multiplicities data.

\end{enumerate}

In this section, we first investigate the sensitivity of each 
parametrization from fits
to the experimental data in the output of interest, namely, the 
hadronic multiplicity
distribution. Then, we discuss in some detail the geometrical 
assumption concerning the multiplicity function and the power law. 
Finally, based on this study, we outline the physical
picture associated with all the results from both 
geometrical/phenomenological and QCD point of views.

\vskip 0.2truecm
\centerline{\bf A. Sensitivity of the parametrizations}
\vskip 0.2truecm

First we observe that the power coefficient $\gamma$ in Eqs. (28-29) 
could, formally, be considered as a free parameter in a direct fit to 
the data
on the hadronic multiplicity distributions and, in this case, it would 
not be necessary to take account of the $e^+ e^-$ average multiplicity 
data. This, however, leds to a
strong correlation between $\gamma$ and the other two inputs, 
$G_{in}(b,s)$
and $\varphi(z)$.
On the other hand, with our procedure, the values and behaviours of the
three inputs, $G_{in}$, $\varphi$ and $\gamma$, are rougly uncorrelated 
and
this allows tests of the inputs by fixing two of them and changing the
 third.
In what follows we perform this kind of analysis, beginning allways 
with
the results obtained in the last section and considering, separately, 
a change in
each one of the inputs.
Since we are interested in the scaling violation we shall base this 
study in
the results for the hadronic multiplicity distribution only at the 
collider energy.

\vskip 0.5truecm

$\bullet$ Changing $G_{in}$

Among the wide class of models for $G_{in}$ \cite{blois}, we shall
consider a multiple diffraction model (MDM) and also the traditional
approach by Chou and Yang, as a class of geometrical model (GM). The 
reason is based on the discussion in Sec. II.A. Also, as we shall 
show in Sec. IV.C, these models allows a suitable connection with the
interpretations that can be inferred from our general approach.

- Multiple diffraction model (MDM)

This class of models is characterized by each particular choice of 
parametrizations for the physical quantities in Eq. (5), namely, form
factors, $G_{A,B}$ and elementary scattering amplitude $f$ \cite{menon}.
In particular, Menon and Pimentel obtained a good description of the 
experimental data o $pp$ and $\overline{p}p$ elastic scattering,
above $\sqrt s = 10\ GeV$ through the following choices 
cite{menonpimentel}:

\begin{equation}
G_{A}(q)=G_{B}(q)= \frac{1}{(1+\frac{q^2}{\alpha ^2})
(1+\frac{q^2}{\beta ^2})},
\end{equation}

\begin{equation}
f(q)=\frac{i[1-(q^2/a^2)]}{[1+(q^2/a^2)^2]}.
\end{equation}

The parameters $a^2$ and $\beta^2$ are fixed and the dependence on the
energy is contained in the other two parameters:

\begin{equation}
C(s)=\xi_{3} exp\{\xi_{4}[{\rm ln}(s)]^{2}\},
\end{equation}

\begin{equation}
\alpha_{(s)}^2=\xi_{1}[{\rm ln}(s)]^{\xi_{2}},
\end{equation}
where $\xi_i$, $i=1,2,3,4$ are real constants. The reason for these 
choices and physical interpretaions are extensively discussed in
\cite{canadian}. With these perametrizations the opacity function,

\begin{equation}
\Omega (b,s) = {\rm Im} \chi (b,s),
\end{equation}
is analytically determined and then the inelastic overlap function
through Eq. (10). 

- Geometrical model (GM)

In the geometrical approach by Chou and Yang, the essential ingredient
is the convolution of form factors in the impact parameter space 
\cite{cy1}. However, in the context of the multiple diffraction theory, 
it can also be specified by the following choices \cite{menon}:

\begin{equation}
G_{A}(q)=G_{B}(q)=\frac{1}{2 \pi [1+ (\frac{q^2}{\mu ^2})]^2},
\end{equation}

\begin{equation}
f=1.
\end{equation}

In Ref. \cite{cy2} the parameters $\mu$ and $C$ were determined 
through fits to elastic $pp$ data at $\sqrt s = 23.5\ GeV$ and 
$\overline{p}p$ at $546\ GeV$. Following the authors, we consider 
the parametrizations

\begin{equation}
C(s) = a_1 + a_2 {\rm ln}s, \qquad {1 \over \mu ^2(s)} = 
b_1 + b_2 {\rm ln}s.
\end{equation}

With the above double pole parametrization for the form factors,
the opacity, Eq. (5), is analytically determined and so the inelastic 
overlap function, Eq. (10).

- Results

The results for the inelastic overlap function at $546\ GeV$ are shown
in Fig. 5, from both the MDM and the GM, together with the BEL $G_{in}$
for comparison. In what follows, we shall use the following notation
for these parametrizations $G^{MDM}_{in}$, $G^{GM}_{in}$, and 
$G^{BEL}_{in}$, respectively.

We then calculate the hadronic multiplicity distribution, Eqs. (28-29), 
at this energy,
by fixing both $\gamma = 0.516$, Eq. (32), and the gamma 
parametrization for
the elementary multiplicity $\varphi(z)$, Eq. (30), and using  
$G^{MDM}_{in}$ and $G^{GM}_{in}$. The results are displayied in 
Fig. 6 togheter with
that obtained with $G^{BEL}_{in}$ (Fig. 4) for comparison.

We observe that, for central collisions (small b), $G^{BEL}_{in}$ and 
$G^{GM}_{in}$ are very similar, but $G^{MDM}_{in}$ has higher values 
(Fig. 5a). This leads to the differences in $\phi(Z')$ at high 
multiplicities, as can be seen in Fig. 6. In the same way, the 
differences between $G^{MDM}_{in}$, $G^{GM}_{in}$, and $G^{BEL}_{in}$
at large b (Fig. 5b) originate the differences in  $\phi(Z')$ at small 
multiplicities. 
In all the cases the physical picture is that large 
multiplicities
(large $Z^{'}$) occur for small impact parameters while grazing 
collisions (large $b$)
lead to small multiplicities, as one would have naively expected.

An important conclusion is that, with $\gamma$ and
$\varphi(z)$ fixed, the hadronic multiplicity distributions obtained
with  $G^{MDM}_{in}$, $G^{GM}_{in}$, and $G^{BEL}_{in}$ reproduce the
experimental data quite well. We shall return to this point in 
Sec. IV.C.

\vskip 0.5truecm

$\bullet$ Changing the elementary distribution $\varphi$

 As a pedagogical exercise, we shall consider only an early
parametrization introduced by Barshay and Yamaguchi \cite{by},

\begin{equation}
\varphi_{BY}(z) = {81\pi^2 \over 64}z^3 exp\{-{9\pi \over 16}z^2\}.
\end{equation}
This function was used in the analysis of $e^+e^-$ multiplicity 
distributions
at lower energies and, as can be seen in Fig 7, does not reproduce 
the
data at higher energies as well as the gamma parametrization.

As before, we now proceed by fixing both $\gamma = 0.516$ and 
$G^{BEL}_{in}$ and using the above parametrization for the elementary
distribution.
The result for the hadronic multiplicity distribution at $546\ GeV$ 
is
shown in Fig. 8, together with the result obtained with the gamma 
parametrization for the elementary process (Fig. 4). The broader 
width of $\varphi_{BY}(z)$ as
compared with that of the gamma distribution, is directly reflected 
in the hadronic multiplicity. Despite the differences between the
two parametrizations for the elementary process the final result for 
the hadronic distribution with $\varphi_{BY}$ can yet be considered 
as a resonable reproduction of the experimental data.

\vskip 0.5truecm

$\bullet$ Changing the power coefficient $\gamma$

 Finally, we consider different parametrizations for the $e^+e^-$
average multiplicity data in the interval $5.1 \leq \sqrt s \leq 183$ 
GeV ,but under the assumption of the power dependence. 
We shall discuss this assumption in the next section.

First we consider the naive parametrization based on the thermodynamic 
model (see next section)

\begin{equation}
<n>_{e^+e^-} = 2.20 [\sqrt s]^{0.500}.
\end{equation}
For the above ensemble of data one obtains $\chi^2/DOF = 209/48 = 4.35$.

Second, and more importantly, we shall investigate the effect of the 
data
at the highest energies, which are not reproduced by our original 
parametrization, as can be seen in Fig. 2. To this end, we consider
only the data above $10$ GeV (25 data points) and the general power 
law parametrization.
With this procedure we obtained

\begin{equation}
<n>_{e^+e^-} = 3.46 [\sqrt s]^{0.396 \pm 0.008}.
\end{equation}
with $\chi^2/DOF = 27/23 = 1.7 $.

The result is displayied in Fig. 9 together with Eq. (42) and our
original parametrization, Eqs. (31-32). We observe that, concerning
$e^+e^-$ average multiplicity Eq. (43) brings information from data
at high energies (roughly above $50$ GeV), while the original
parametrization, Eqs. (31-32) is in agreement with data at smaller 
energies
(below $\sim 100$ GeV) and the same is true for the parametrization 
with Eq. (42).

As before, we now calculate the corresponding hadronic multiplicity 
distribution by fixing both the gamma parametrization for the 
elementary distribution, Eq. (30), and the $G^{BEL}_{in}$, Eqs. 
(22-24), and considering the three parametrizations for the average 
multiplicity, Eqs. (31-32), (42) and (43).
The results at $546\ GeV$ are shown in Fig. 10.

We conclude that, in the context of our approach with the fixed 
inputs 
$G^{BEL}_{in}$ and gamma parametrization for $\varphi(z)$, the 
information from the $e^+e^-$ average multiplicites at high energies
with the power-law does not reproduce the hadronic multiplicity 
distribution. That is,
the  elementary average multiplicity distributions in hadronic 
interactions must deviates from the $<n>_{e^+e^-}(s)$ as the energy
increases, roughly above $\sim 50 - 100\ GeV$. We shall discuss 
the physical interpretations of this result in Sec. IV.C.

\vskip 0.2truecm
\centerline{\bf B. The multiplicity function and the power assumption}
\vskip 0.2truecm

We now turn to the discussion of a crucial assumption in our approach, 
namely, that the {\it elementary} average multiplicity at fixed impact
 parameter collisions grows as a power of the center-of-mass energy. 
To this end we shall first briefly recall some aspects of the 
power-law in hadron-hadron and $e^+e^-$ collisions, both in experiment 
and theory, and after, based on these ideas, we shall present a 
discussion concerning the use of this assumption in our approach 
and also the meaning of the multiplicity function.

From the early sixties cosmic ray results on extensive air showers, 
at energies
$E_{lab} < 10^6 - 10^7 \ GeV$, led to empirical fits of the type 
$<N> \propto E_{lab}^{1/4} \propto [\sqrt s]^{1/2}$ 
(see \cite{feinberg} for a review). A general power law with the 
exponent as a free parameter was used a long time ago, in order to 
allows analytical connections in analysis of cosmic
ray data \cite{shibuya}. Also, in the beginnig of sixties, these 
investigations introduced the concept of inelasticity \cite{feinberg}. 
This comes from the observation that the energy effectively available 
for particle production could not be identified with the c.m. energy, 
as believed before $1953$ (Wataghin, Fermi, Landau), but only with a 
fraction of it:

\begin{equation}
W = k \sqrt s .
\end{equation}
The remained $(1 - k)\sqrt s$ was associated with the early named
 ``isobar'' system,
presently known as leading particle.

From the theoretical side, the power dependence emerged in the context 
of statistical models (Fermi, Pomeranchuck) and hydrodynamical models 
(Heisenberg, Ladau) \cite{feinberg}. For example, taking account of 
the inelasticity, in the Landau model, the fact that the averaged 
multiplicity is proportional to the total entropy leads to the result 
\cite{wilk}
\begin{equation}
<N> \propto k^{3/4} [\sqrt s]^{1/2}.
\end{equation}
Dependences on $s^{1/2}$ is characteristic of the Heisenberg and 
Pomeranchuck models and even $s^{1/8}$ appears in the Landau model, 
when viscosity is taken into account 
\cite{feinberg}. In the context of termodynamic models, a universal 
formula was discovered for proton targets and for energies below 
$\sim 50\ GeV$: Data including $\gamma, \pi, N$ and $p$ collisions 
with $p$ were quite well reproduced by $<N> = 1.75\ s^{1/4}$ 
\cite{carruthers}.
Concerning $e^+e^-$ data on average multiplicity this model 
suggested $<n> = 1.5 \ s^{3/8}$ and pure fits to low energy data 
furnished 
$<n> = (2.2 \pm 0.1) s^{0.25\pm 0.01}$, and also  $<n> = 
(1.73 \pm 0.03) s^{0.34\pm 0.01}$ \cite{criegee,berger}. Moreover, 
the power-law, with the exponent $1/4$, was successfully used in the
 context of the parton model, either connecting KNO and Bjorken 
scaling \cite{eilam} or treating directly the violation of the KNO 
scaling \cite{rudaz}.

The power-law may also appear under more general arguments. For 
example, suppose that an intermediate state (fireball) of invariant 
mass ${\sl M} \propto \sqrt s $ decays into two systems each of 
invariant mass ${\sl M}_1 = {\sl M} / c$, where $c$ is a constant. 
Suppose also that similar processes continue through some steps 
(sucessive cluster production) until the masses reach a value 
${\sl M}_0$ (some minimum ressonance mass). It is easy to show 
that the final multiplicity reads \cite{rudaz}
\begin{equation}
n \propto [\sqrt s]^{{\rm ln} 2/ {\rm ln} c}.
\end{equation}
For example for $c=4 \rightarrow n \propto s^{1/4}$. The exponent 
$\gamma$ (our notation) may be inferred from $c = 2^{1/\gamma}$, 
so that higher $\gamma$ values
imply in higher splited masses in each step (for $\gamma = 0.516 
\rightarrow c \sim 3.8$).

Based on the above review, we see that the power-law is 
characteristic of several analysis of experimental data on 
hadron-hadron and $e^+e^-$ collisions and also several theoretical 
approaches and models. Now we shall discuss this law in the context 
of our approach.

First let us stress that in our formalism this assumption concerns 
an {\it elementary} hadronic process taking place at {\it fixed 
impact parameter b}. Thus, it does not pretend to represent the 
average hadronic multiplicity $<N(s)>$. Also, we used $e^+e^-$ 
data only as a possible source of information on contact 
interactions (fixed $b$) and therefore the power assumption does 
not pretend to represent the average multiplicity in $e^+e^-$ 
collisions. This is a subtle point in our approach and we would 
like to discuss it in some detail.

The main reason for the power assumption was to obtain an analytical 
and simple connection between the multiplicity function $m(b,s)$ and 
the eikonal, Eq. (27), which allows the general analytical connection 
between the elastic and inelastic channels. Since it
is typical of several kind of collisions, as reviewed above, it is not 
unreazonable that it could represent an elementary hadronic process 
taking place at fixed impact parameter. Just for its elementary 
character 
(at given $b$), there seems also to be no reason to include any 
inelasticity effect (leading particle) in the basic assumption 
represented by Eq. (25). That is, it seems reasonable that $<n>(b\ 
fixed, s)$ may be just proportional only to $E_{CM}^{\gamma}$.

The multiplicity function $m(b,s)$, as defined by Eq. (20), connects 
the hadronic and elementary (at given $b$) average multiplicities. 
With the power assumption and the geometrical arguments of Sec. 
II.C, $m(b,s)$ may be expressed in terms of the eikonal
and the power coefficient $\gamma$. The subtle point in our approach 
is that, since by definition $m(b,s)$ is proportional also to the 
average elementary multiplicity at given $b$, the coefficient 
$\gamma$ was determined by fit to data available on contact 
interactions. In this sense, the model ``imposes'' the power-law 
and the $e^+e^-$ data  are supposed to provide the limited, but 
possible, information on contact interactions.

These considerations may allow to infer a distinction between 
$e^+e^-$ average
multiplicity and what this quantity could be in an elementary 
hadronic process.
Specifically, we showed in Sec. IV.A that data on the average 
multiplicity in $e^+e^-$ collisions, presently available above 
$5\ GeV$,
can not be reproduced by the power-law. For example, a second 
degree polynomial in $ln s$ gives a quite good fit to all the
data above $5\ GeV$:

\begin{equation}
<n>_{e^+e^-}(s) = 0.0434 + 0.775{\rm ln}s + 0.168{\rm ln}^2s
\end{equation}
with $\chi^2/DOF = 145 / 45 = 3.2$. However, besides this 
parametrization does not allow the analytical connection with 
the eikonal, we showed that, with the power-law, the
behavior of $<n>_{e^+e^-}(s)$ at energies above $\sim 50\ GeV$ 
does not
lead to the description of the hadronic multiplicity distribution. 
In
other words, in the context of our approach, the increase of the 
elementary average multiplicity with energy in hadronic collisions
 must be faster than that observed in $e^+e^-$ collisions. This 
is not the case at lower energies, since the power-law with
 $\gamma = 0.516$ gives a satisfactory description of the $e^+e^-$
 data. In the next section we discuss the physical interpretations
 associated with these observations.

\vskip 0.2truecm
\centerline{\bf C. Physical picture}
\vskip 0.2truecm

Based on the results of Secs. II and III, we now discuss the physical
picture associated with the scaling violations, specifically, with the
evolution of the hadronic multiplicity distribution $\Phi(Z')$ from
the ISR to the collider and LHC energies, Figs. 3 and 4.

From Eq. (21) the hadronic multiplicity $\Phi$ is constructed in 
terms of $G_{in}$ and the elementary
quantities $\varphi$ and $m$. In our approach, $\varphi$ scales and
so does not depend on the energy. The multiplicity function $m(b,s)$
is connected with $G_{in}$ through Eqs. (10) and (27),

\begin{equation}
m(b,s) = \xi(s)\{ {\rm ln}[ 1 - G_{in}(s,b)]\}^{\gamma},
\end{equation}
where $\xi$ comes from the normalization condition (29). Both
$\xi(s)$ and $m(b,s)$ depend on the power coefficient $\gamma$,
which is a constant determined from the fit through Eq. (31).
Therefore, the evolution of the hadronic distribution with energy 
comes
directly from $G_{in}(b,s)$ and depends also on the value of the
exponent $\gamma$. This exponent, in turns, comes from the elementary
average multiplicity dependence with the energy, Eq. (31), and 
therefore is associated with the {\it effective} number of colliding 
constituents in the hadronic process.

Based on the above observations, the physical picture that emerges 
is that the energy evolution of the hadronic multiplicity distribution 
is correctly reproduced by changing only the overlap function, without
tampering with the underlying more elementary process ($\varphi $). 
The geometrical evolution of the constituents of the hadron is 
responsible for the energy dependence and not the dynamical interaction 
itself. This
is what one would expect if the underlying interaction is unique (QCD)
but the relative importance of the constituents involved in collisions 
changes with energy (indicated by the exponent $\gamma$).

We showed that with the power assumption, the information from $e^+e^-$ 
data above,
say, $\sim 100$ GeV leads to an understimation of the hadronic 
multiplicity ditribution (Fig. 10). This means that the average 
multiplicity in an elementary hadronic process must increase with 
energy faster than that associated with $e^+e^-$ collisions.
This result seems quite reasonable 
since, in a QCD guided approach, we expect different contributions 
from gluons/quarks interactions 
than those associated with lepton-lepton collisions. As the average 
multiplicity increases, the relevance of the original parton decreases, 
so that at high energies $e^+e^-$ can serve as a good first guide to 
quark-quark, quark-gluon and gluon-gluon multiplicity distributions.
In a parton model (following QCD), this effect above $\sim 100$ GeV 
may be interpreted as the unset of gluons interactions \cite{rudaz}. 
The faster increase represented by our power-law with $\gamma = 0.516$ 
(Fig. 2) may be atributed to the
fully development of the gluonic structure, rather than the quark 
(valence) structure.

These ``microscopic'' interpretations are also directly associated
 with the BEL behavior, since its origin may be traced either to gluon 
interactions in the eikonal formalism \cite{margolis1,margolis2}, or
 to the increased size of spot scattering in the overlap function 
formalism \cite{henzi}.

As commented before, a novel aspect of this work concerns the 
simultaneous treatment of both the elastic and inelastic channels. 
Specifically, we started from elastic channel descriptions ($pp$ and 
$\overline{p}p$ differential cross sections) and extended the results 
to the inelastic channel (multiplicity distributions). In this sense, 
we expect that the physical picture from both channels should be 
the same.
Besides the microscopic interpretation associated with the BEL 
$G_{in}(b,s)$, even if we consider the naive models represented by 
the MDM and the GM, discussed in Sec. IV.A, the same scenario 
emerges. In fact, in both models the elementary interaction, 
represented by the elementary elastic amplitude $f(q,s)$, does 
not depend on the energy, Eqs. (34) and (39). The energy dependence 
is associated with the form factor $G(q,s)$ and the ``absorption 
constant'' $C(s)$. The former, through the associated radius,
\begin{equation}
R^2(s) = -6 {dG \over dq^2} \vert _{q^2 = 0},
\end{equation}
describes the expansion effect (geometry). The latter is associated 
with absorption (blackning) in the context of the geometrical 
(Chou-Yang) model and to the relevant number on constituents in 
the context of the multiple diffraction theory \cite{glauber}. 
herefore, we also conclude that the elementary interaction is unique
 (does not depend on the energy), but the geometrical evolution of 
the constituents and its
relevent number in collisions changes as the energy increases.

\vskip 0.3truecm

\centerline{\bf V. CONCLUSIONS AND FINAL REMARKS}

\vskip 0.3truecm

The underlying theory of the hadronic phenomena is QCD. As commented 
in Secs. I and II, depite all its successes, the theory has presently 
some limited efficiency in the treatment of soft hadronic processes,
 meanly related with unified descriptions of physical quantities from 
{\it both} elastic and inelastic channels. Moreover, some QCD 
approaches are based in extensive Monte Carlo calculations and 
concerning this point, we understand that, although these techniques 
represent a powerfull tool for experimentalists, it is questionable 
if they could really be the adequate and final scenario for a 
theoretical understand of the hadronic interactions, mainly if we 
think in connections with first principles of QCD.

At this stage, it seems that phenomenology must play an important 
role to bridge the gap, or, at least, to indicate or suggest some 
suitable calculational schemes for further theoretical developments. 
On the other hand, all the phenomenological approaches presently 
available, have also very limited intervals of validity and efficiency 
in the treatment of hadronic processes at high energies. One of the 
serious limitations of the geometrical approach is the difficulty to 
directly connect its relative efficiency with the well established 
microscopic ideas (QCD). However, it has not been proved that this
 direct connection can not be obtained.

In this work, making use of the unitarity principle and in the context 
of a geometrical picture, we obtained analytical connections between 
physical quantities from both elastic and inelastic channels. In 
particular we correlated quantitatively the violations of the geometrical 
and KNO scalings in an analytical way. The physical picture that emerges
 from both channels, for $pp$ and $\overline{p}p$ collisions above 
$\sim 10\ GeV$ is the following. The dependence of the physical 
quantities with the energy (elastic differential cross section and 
inelastic multiplicity distributions) is associated with the geometrical 
evolution of the constituents and the relative importance of the 
constituents involved in the collisions. The underlying elementary 
process or interaction does not change with the energy. This is in 
agreement with what could be expected from QCD.

With this kind of approach the correct information extracted from the 
elastic
channel is fundamental. Our prediction at LHC energies was based on
extrapolations from analysis at lower energies and so has a limited 
character. This observation, and obviously other considerations 
regarding different models, point out to the importance of complete 
measurements 
of physical quantities associated with the elastic channel at the 
LHC,
that is not only total cross sections but also the $\rho$ parameter 
and
differential cross sections at large momentum transfer.

Based on the limitations referred to in this section, we 
do not pretend that the forms we inferred for the hadronic 
constituent-constituent collisions, multiplicity distribution 
and average multiplicity, are a conclusive solution. However, we 
hope that, at least, they can bring new information on what some 
aspects of an 
elementary hadronic process could be.

\vskip 0.5truecm

\centerline{\bf ACKNOWLEDGEMENTS}
Thanks are due to Pierre St. Hilaire. P.C.B. and M.J.M. are 
thankful to CNPq and FAPESP ( Proc. N. 1998/2249-4) for financial 
support.

\newpage

{\bf Figure Captions}

\vs{1.0cm}

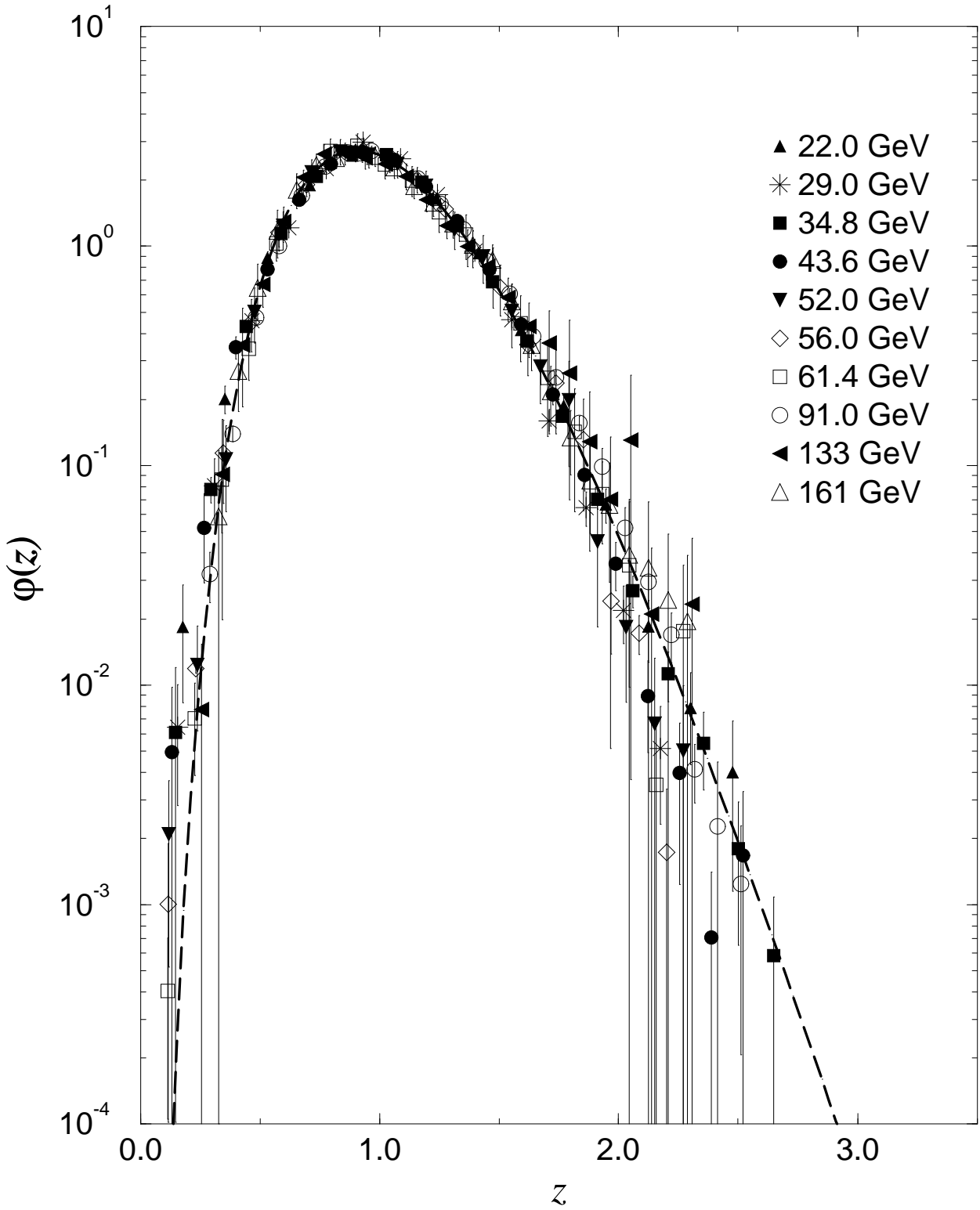
Fig. 1. The KNO charged multiplicity distribution for $e^{+}e^{-}$ 
annihilation data \cite{delphi,data1} and the fitted gamma distribution, Eq. (30) 
(dashed).

\vs{0.5cm}

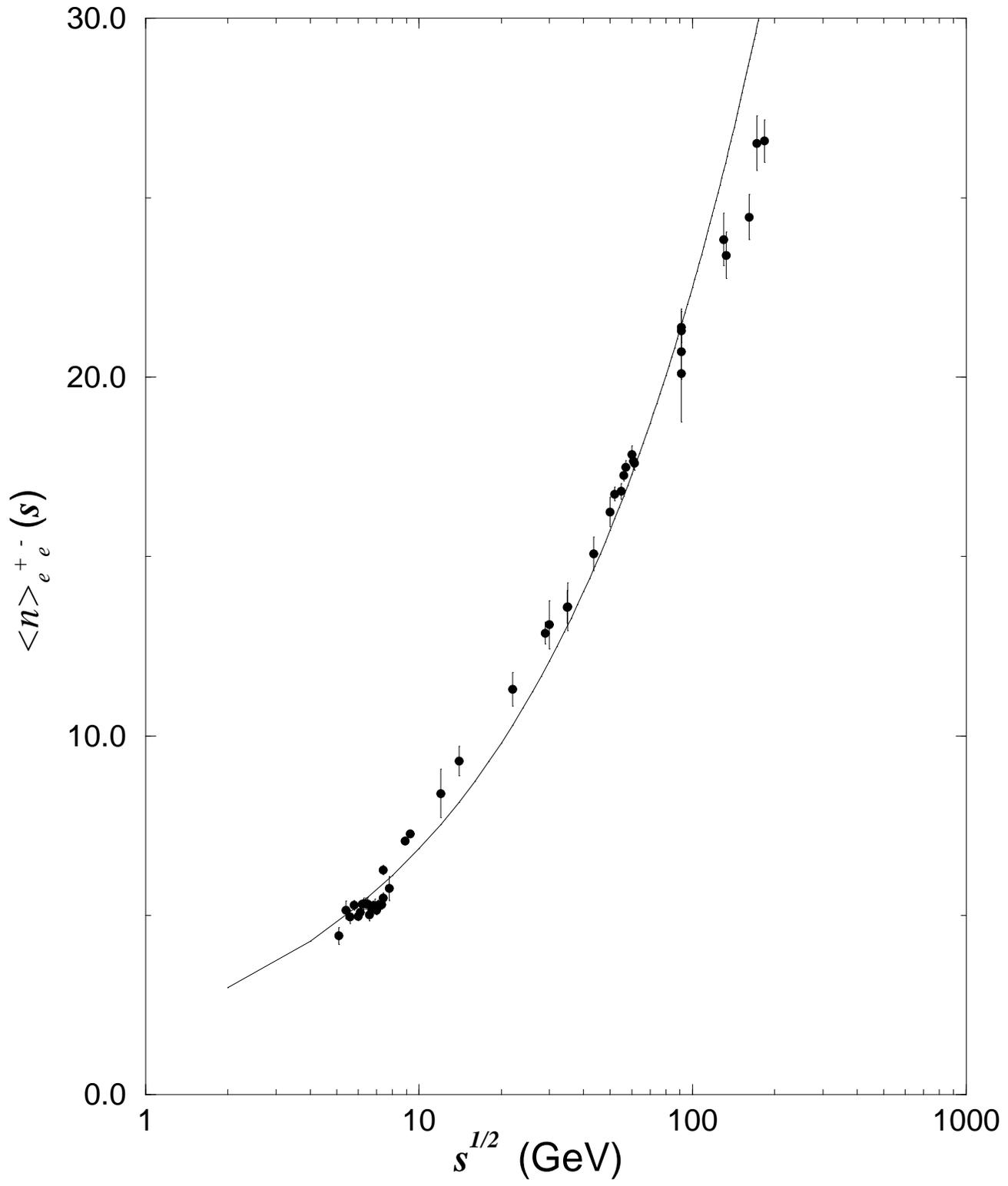
Fig. 2. The average charged multiplicity for $e^{+}e^{-}$ annihilation
 data \cite{data1,data2} and the fitted power law, Eq. (31). 

\vs{0.5cm}

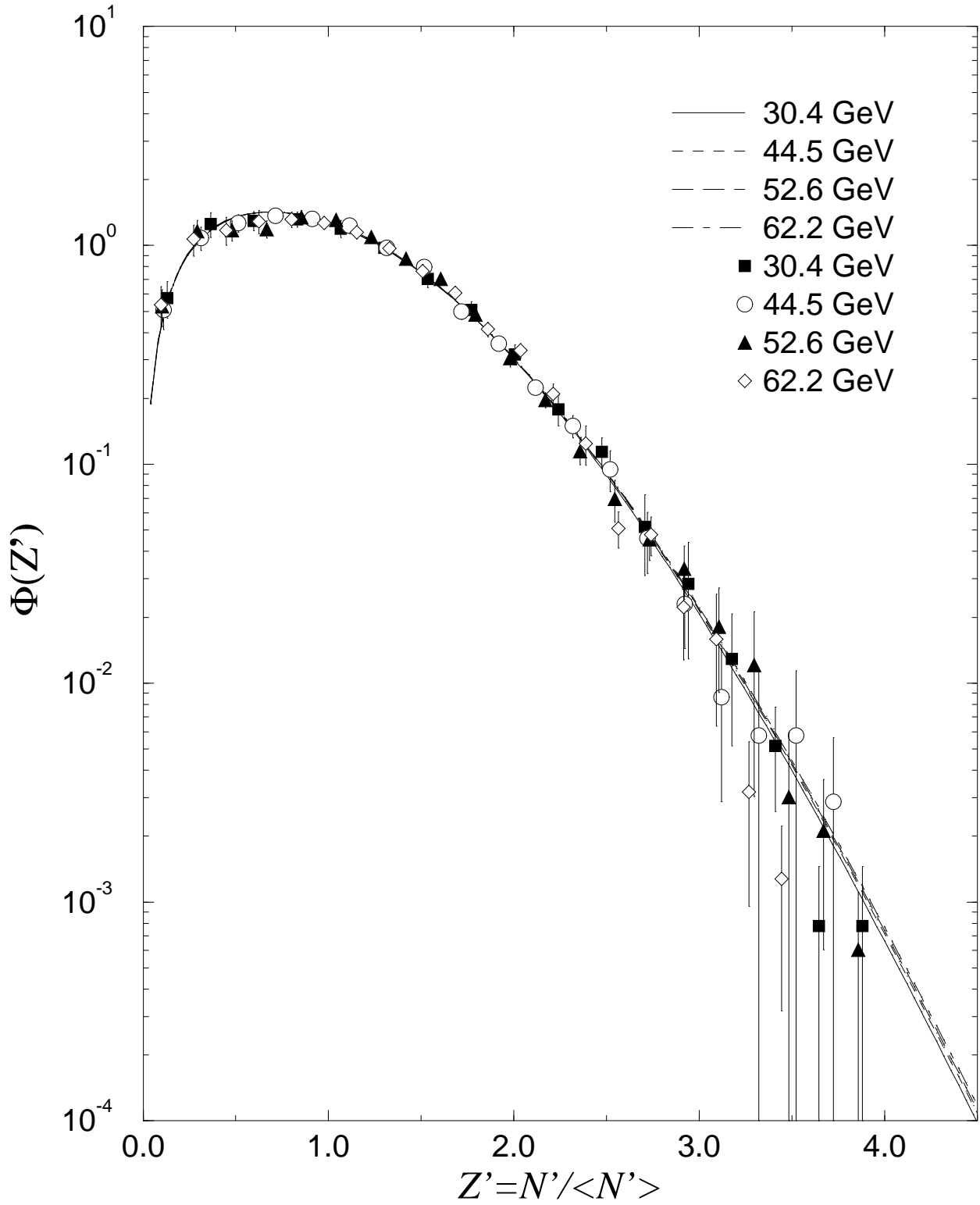
Fig. 3. Scaled multiplicity distribution for inelastic $pp$ data 
\cite{break} 
at ISR energies compared to theoretical expectations using Eqs. 
(28-29).

\vs{0.5cm}

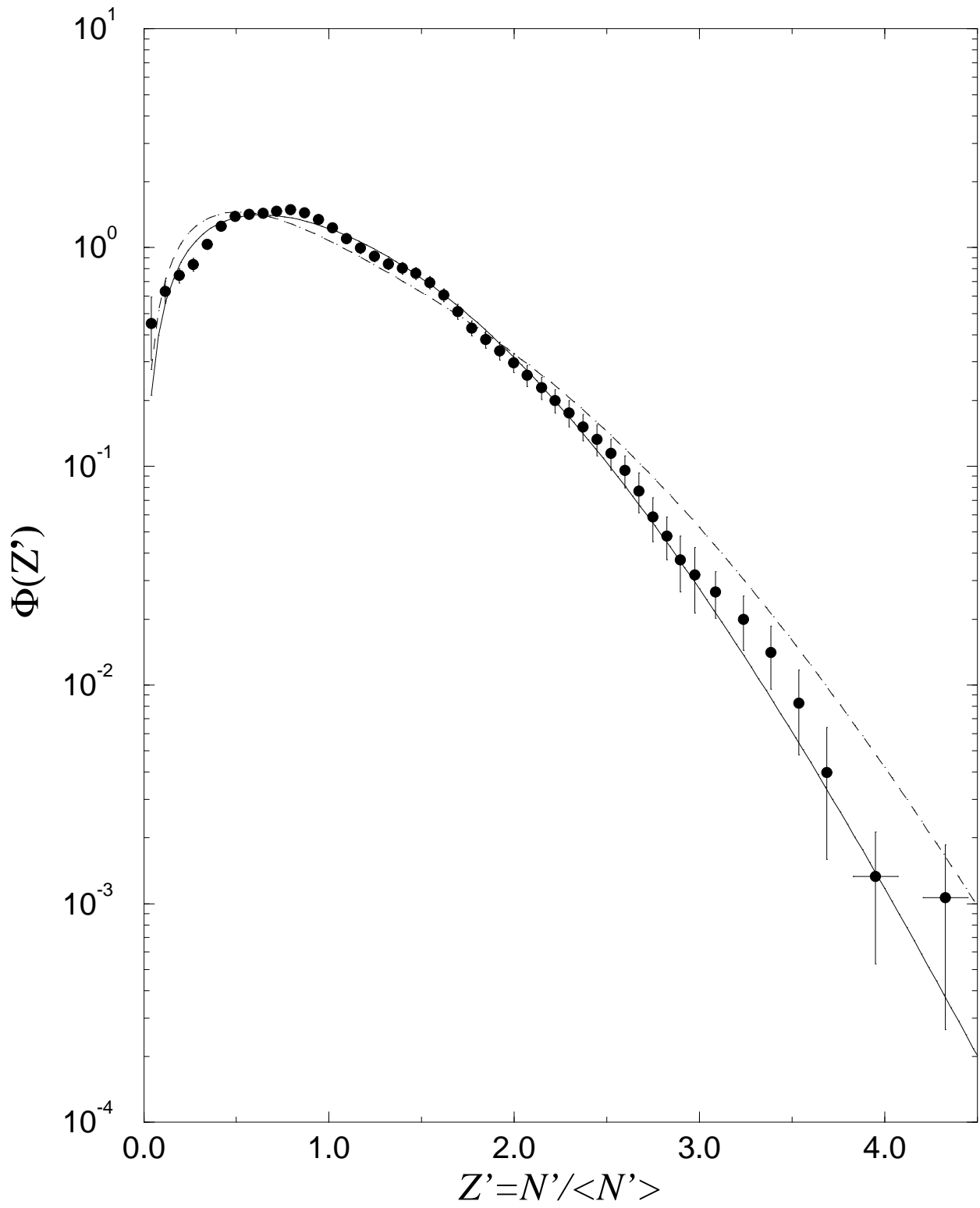
Fig. 4. Scaled multiplicity distribution for inelastic 
$\overline{p}p$ data \cite{ua5} at 546 GeV  compared to theoretical 
expectations using Eqs. (28-29) (solid) and predictions at 14 
TeV (dashed).

\vs{0.5cm}

Fig. 5. Inelastic overlap functions for $\overline{p}p$ collisions 
at $546\ GeV$, predicted by the multiple diffraction model (MDM), 
geometrical model (GM) and the short-range-expansion, 
black-edge-large approach (BEL): (a) central region; (b) large 
distances.

\vs{0.5cm}

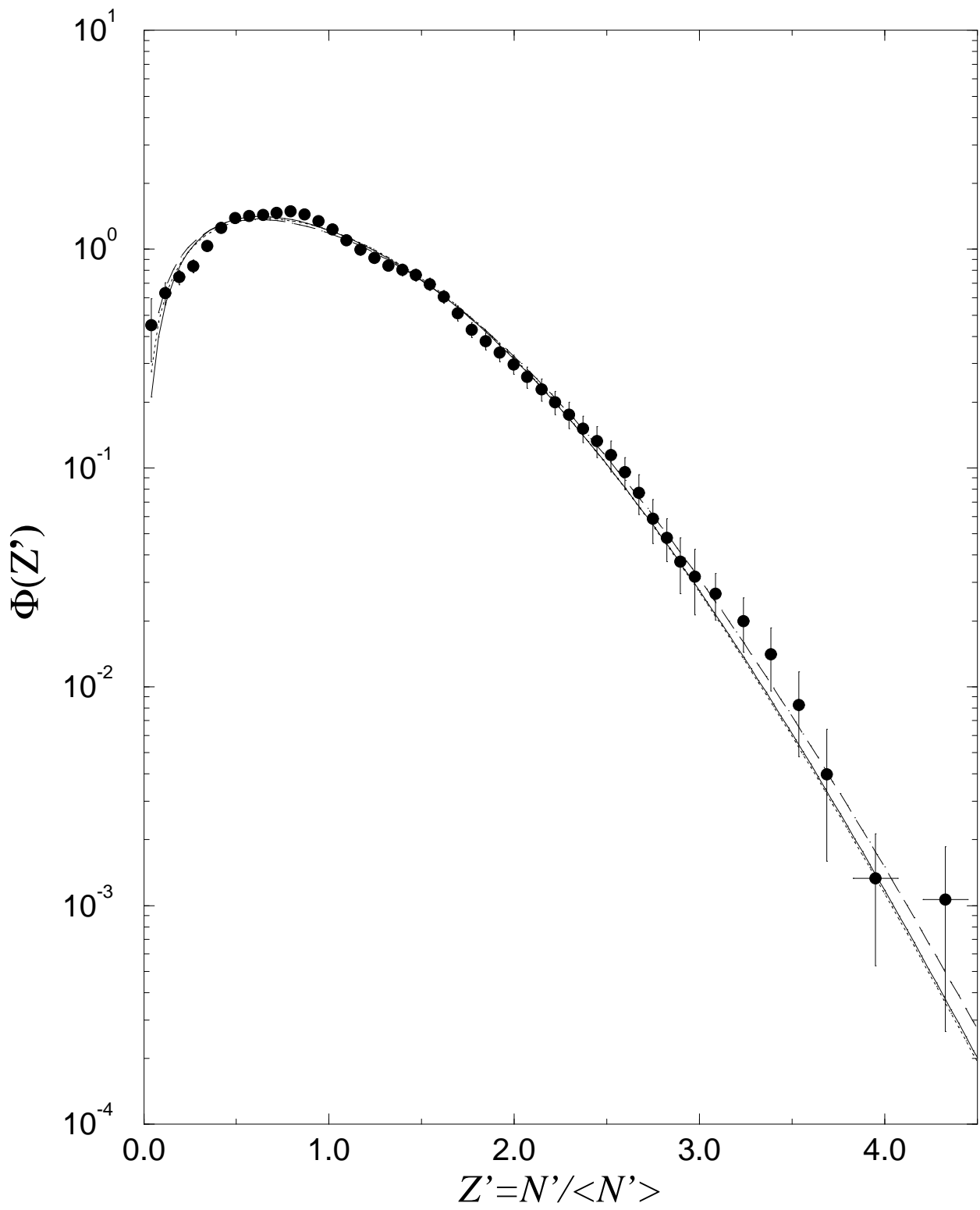
Fig. 6. Same as Fig. 4 with the three different inputs for the 
inelastic overlap function. Same legend of Fig. 5.

\vs{0.5cm}

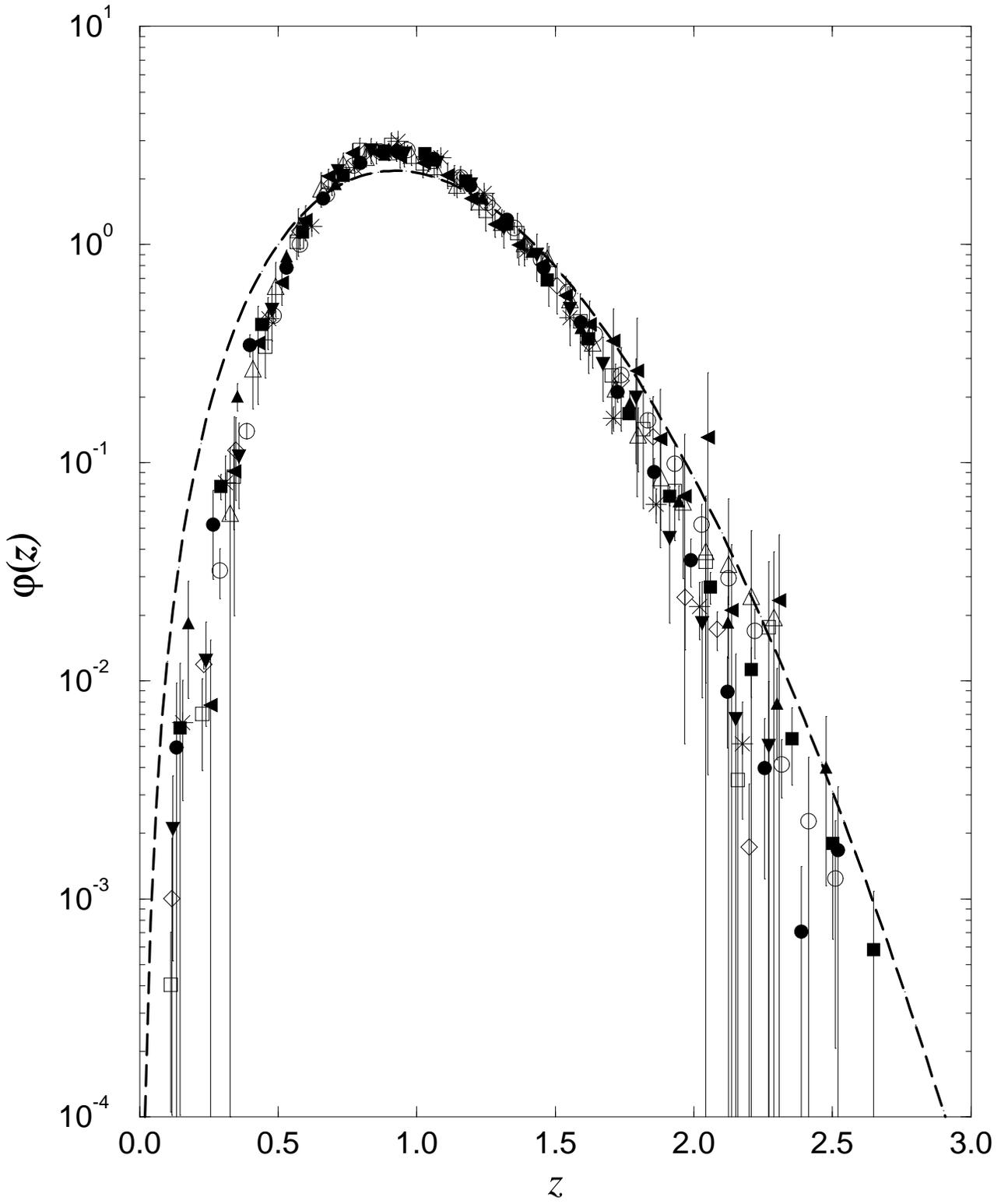
Fig. 7. Same as Fig. 1 with the Barshay-Yamaguchi parametrization, 
Eq. (41).

\vs{0.5cm}

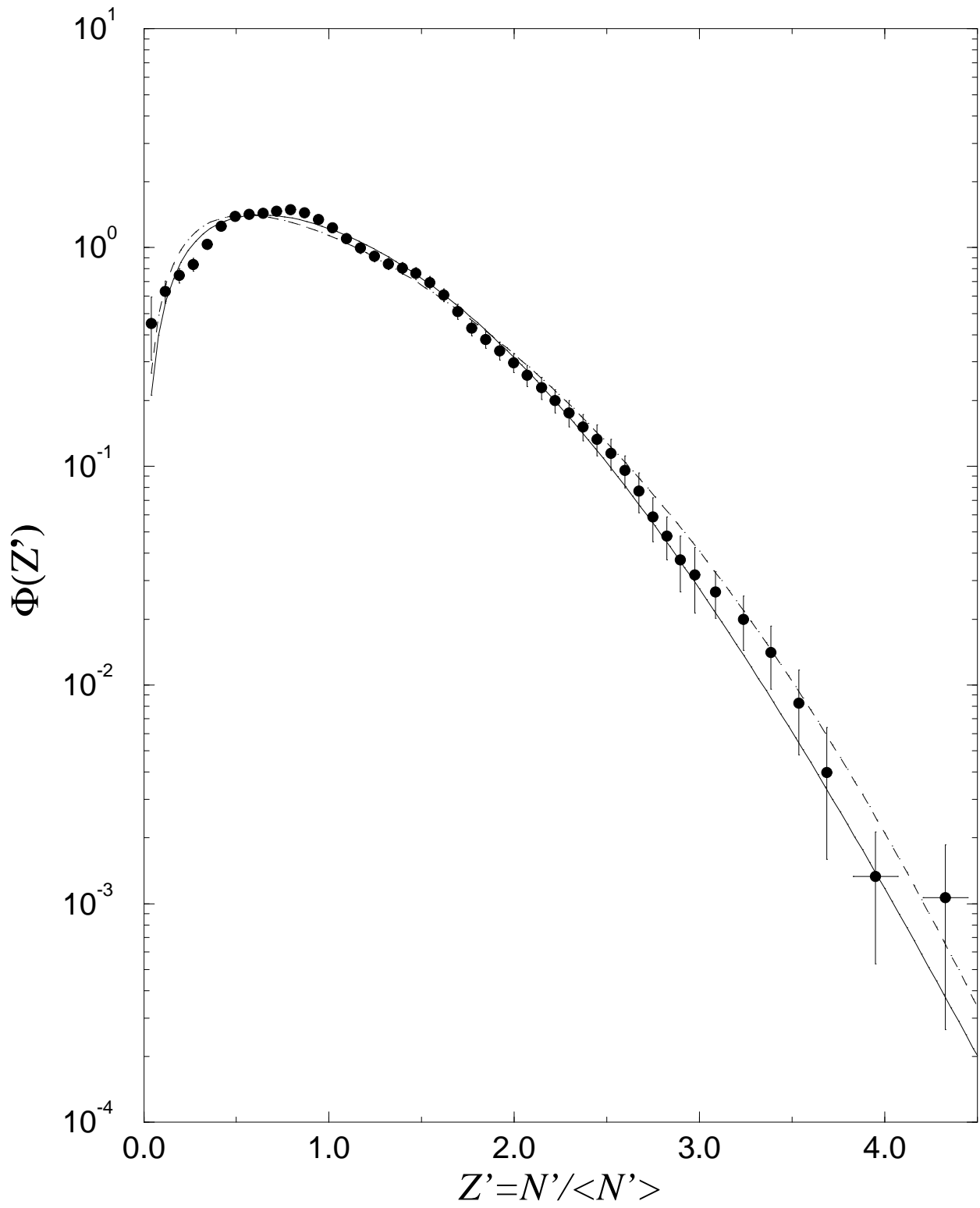
Fig. 8. Same as Fig. 4 with two different inputs for the elementary 
multiplicity distributions: gamma function, Eq. (30) (solid) and 
Barshay-Yamaguch parametrization, Eq. (41) (dot-dashed).

\vs{0.5cm}

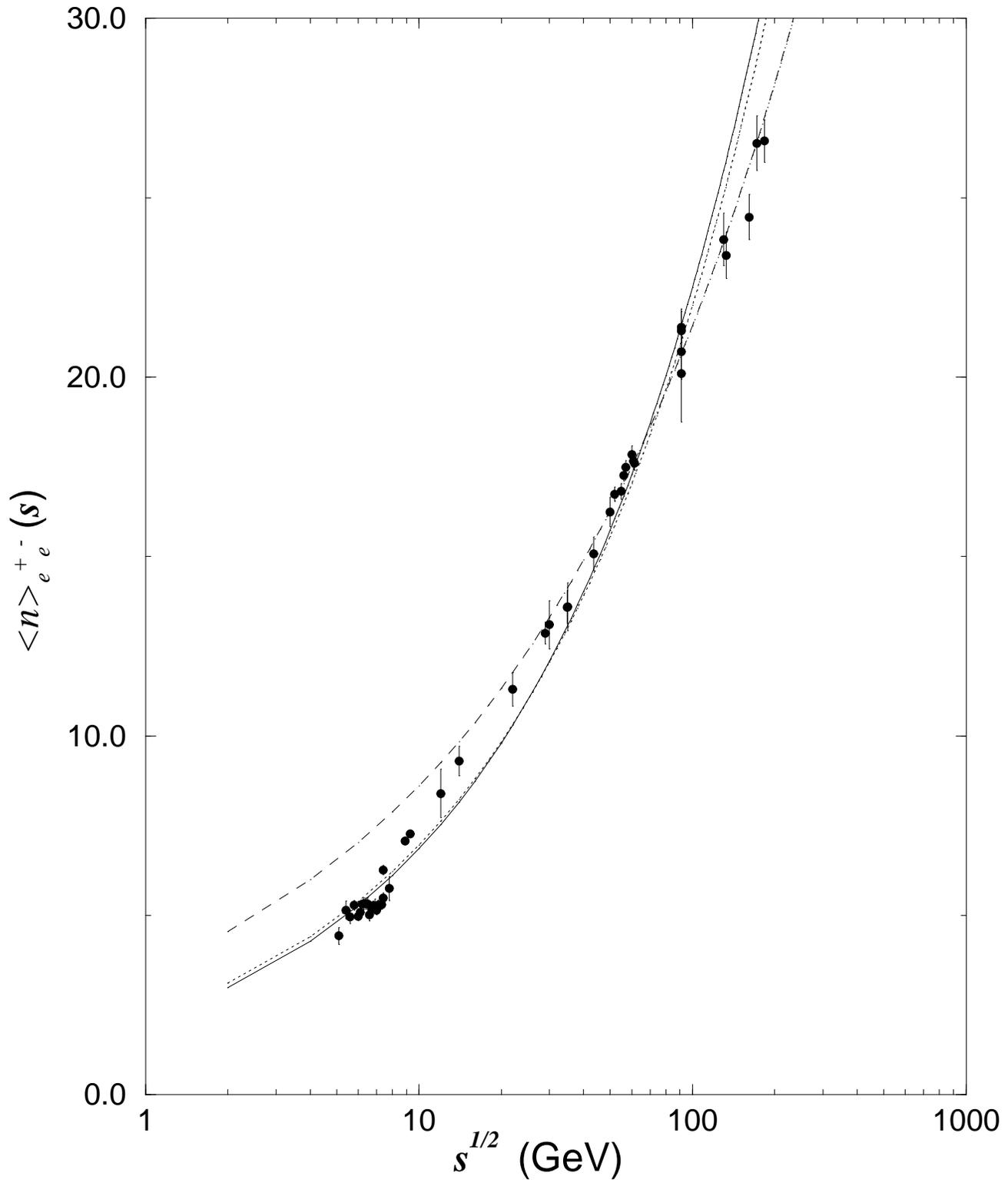
Fig. 9. Same as Fig. 2 with three power-law parametrizations: 
$\gamma = 0.516$ (solid), $\gamma = 0.500$ (dotted) and $\gamma = 
0.396$ (dot-dashed). In the last case only data above $10\ GeV$ 
were fitted.

\vs{0.5cm}

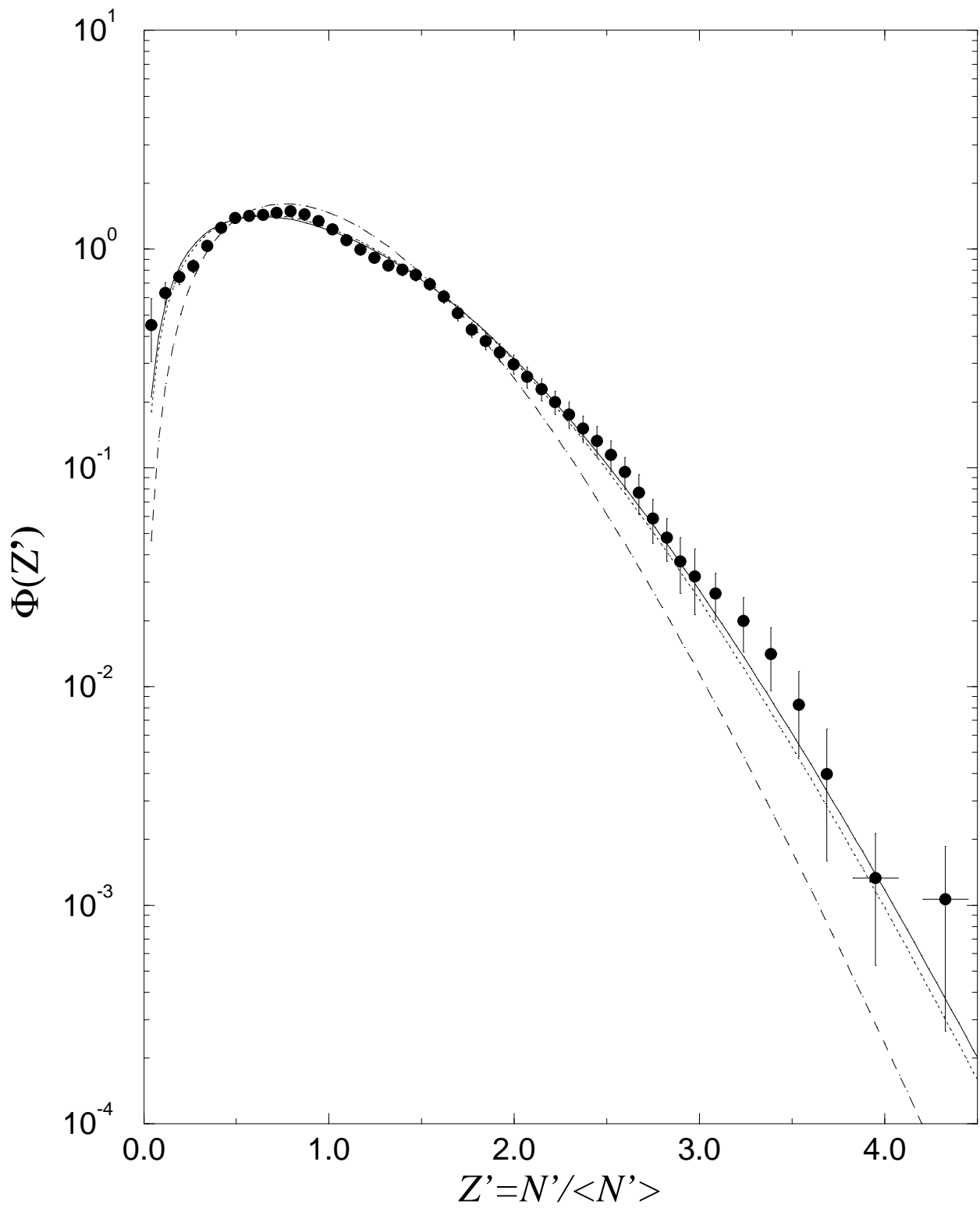
Fig. 10. Same as Fig. 4 using the three different parametrizations 
for the elementary
average multiplicity (Fig. 9 and same legend).

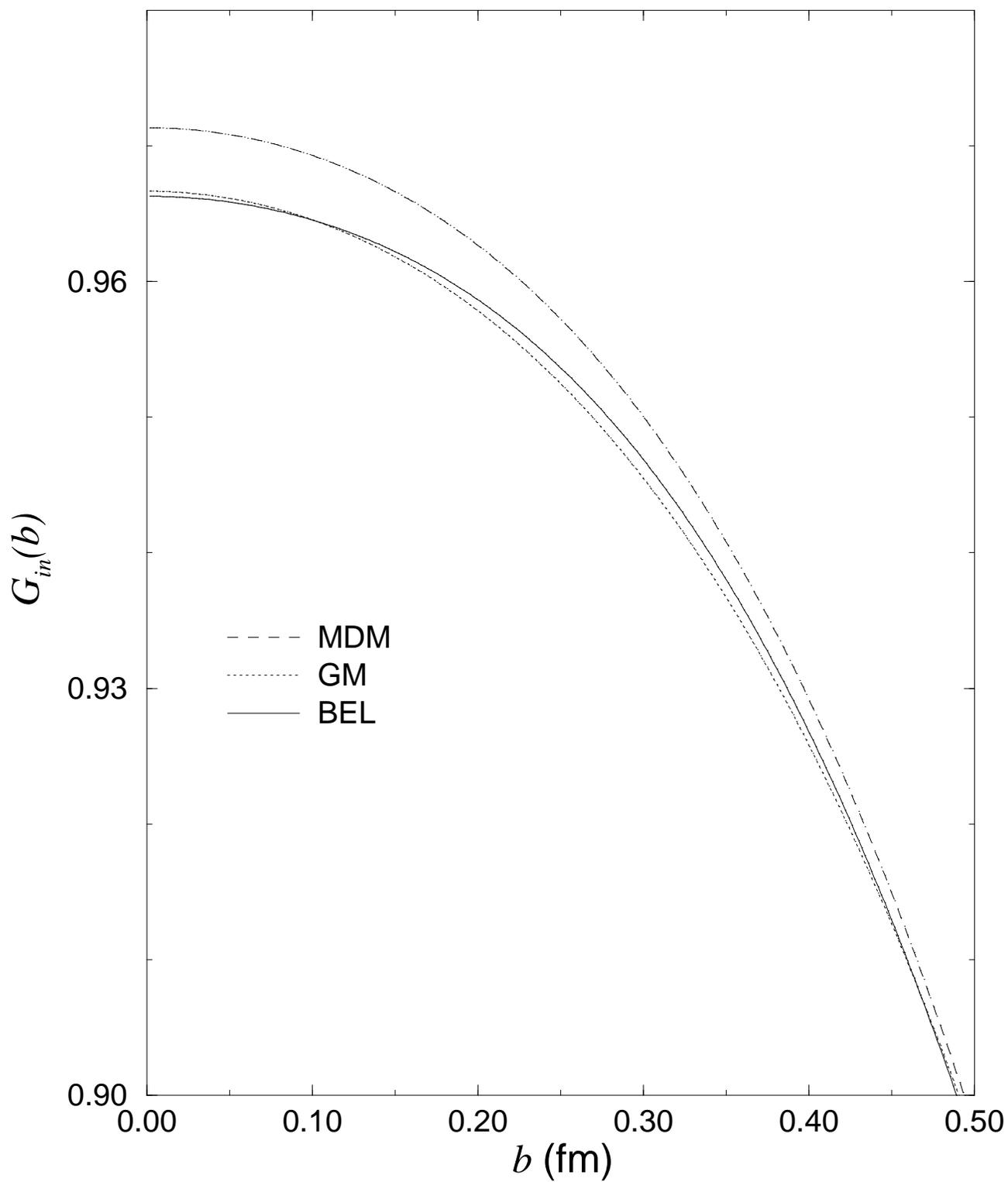
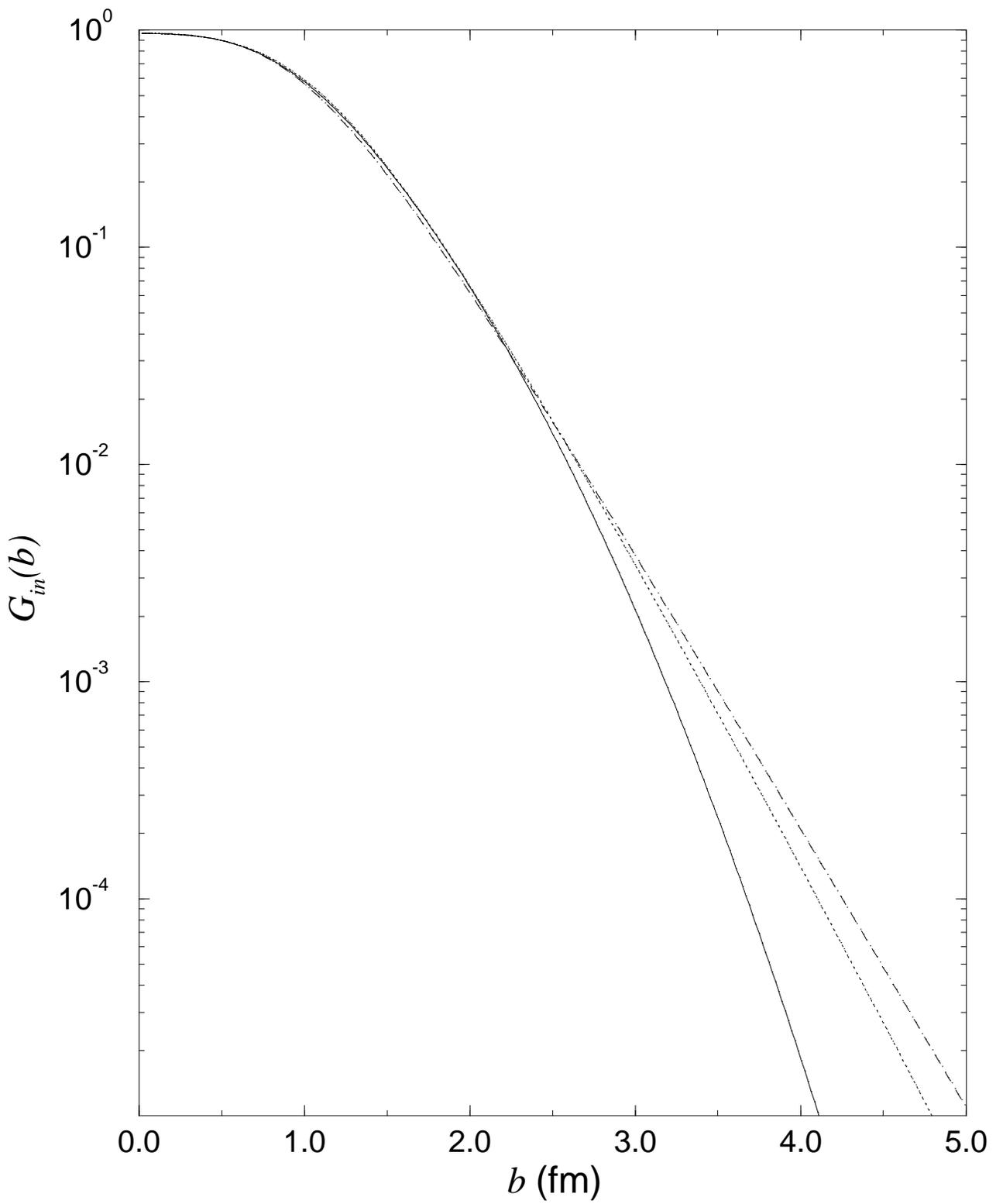
\end{document}